\RequirePackage{lineno}
\documentclass[prd,aps,noeprint,superscriptaddress,twocolumn,amsmath,amssymb]{revtex4-2}
\usepackage[colorlinks,linkcolor=blue,anchorcolor=blue,citecolor=blue]{hyperref}
\usepackage{lineno}
\usepackage{graphicx}
\usepackage{subfigure}
\usepackage{bm}
\usepackage{overpic}
\usepackage{verbatim}
\usepackage{rotating}
\usepackage{tabularx}
\usepackage{array}
\usepackage{xparse}

\usepackage{dcolumn}
\newcolumntype{C}[1]{>{\centering}p{#1}}
\newcolumntype{L}[1]{>{\raggedright}p{#1}}
\newcolumntype{R}[1]{>{\raggedleft}p{#1}}
\newcolumntype{P}{@{$\ \pm\ $}}

\newcommand{\unitegev}{~{\rm GeV}}
\newcommand{\unitemev}{~{\rm MeV}}

\newcommand{\unitmgev}{~{\rm GeV}/c^2}
\newcommand{\unitmmev}{~{\rm MeV}/c^2}
\newcommand{\unitenergyg}{~{\rm GeV}/c^2}

\newcommand{\ra}{\rightarrow}
\newcommand{\ee}{e^+e^-}
\newcommand{\pip}{\pi^+}
\newcommand{\pim}{\pi^-}
\newcommand{\piz}{\pi^0}

\newcommand{\chisq}{\chi^{2}}
\newcommand{\chisqfour}{\chi^{2}_{\rm 4C}}

\newcommand{\luminosity}{647~{\rm pb}^{-1}}
\newcommand{\energy}[1][2.1250]{\sqrt{s} = #1 \unitegev}

\newcommand{\opp}{\omega\piz\piz}
\newcommand{\opm}{\omega\pip\pim}
\newcommand{\ppp}{\pip\pim\piz}
\newcommand{\pipipi}{\ppp}
\newcommand{\processopp}{\ee\ra\opp}
\newcommand{\processopm}{\ee\ra\opm}

\newcommand{\pizone}{\piz_{1}}
\newcommand{\piztwo}{\piz_{2}}
\newcommand{\pizthree}{\piz_{3}}
\newcommand{\pizonefull}{\gamma_{1} \gamma_{2}}
\newcommand{\piztwofull}{\gamma_{3} \gamma_{4}}
\newcommand{\pizthreefull}{\gamma_{5} \gamma_{6}}

\newcommand{\pipipionefull}{\pip\pim\pizonefull}

\def\babar{\mbox{\slshape B\kern-0.1em{\small A}\kern-0.1em B\kern-0.1em{\small A\kern-0.2em R}}}

\newcommand{\resultoppmass}{2222}
\newcommand{\resultoppmasserrsta}{7}
\newcommand{\resultoppmasserrsys}{2}
\newcommand{\resultoppwidth}{59}
\newcommand{\resultoppwidtherrsta}{30}
\newcommand{\resultoppwidtherrsys}{6}
\newcommand{\resultoppsignificance}{5.3}
\newcommand{\resultoppcodemass}{$M=\resultoppmass\pm\resultoppmasserrsta\pm\resultoppmasserrsys\unitmmev$}
\newcommand{\resultoppcodewidth}{$\varGamma=\resultoppwidth\pm\resultoppwidtherrsta\pm\resultoppwidtherrsys\unitemev$}
\newcommand{\resultoppcodesignificance}{$\resultoppsignificance\sigma$}

\begin{document}
\normalsize
\parskip=5pt plus 1pt minus 1pt
\title{\boldmath Measurement of the $\processopp$ cross section at center-of-mass energies
    \\from 2.0 to 3.08 GeV}

\author{
    M.~Ablikim$^{1}$, M.~N.~Achasov$^{10,b}$, P.~Adlarson$^{67}$, M.~Albrecht$^{4}$, R.~Aliberti$^{28}$, A.~Amoroso$^{66A,66C}$, M.~R.~An$^{32}$, Q.~An$^{63,50}$, X.~H.~Bai$^{58}$, Y.~Bai$^{49}$
% ...
Yi2020~Wang$^{53}$, Ying~Wang$^{51}$, Z.~Wang$^{1,50}$, Z.~Y.~Wang$^{1,55}$, Ziyi~Wang$^{55}$, D.~H.~Wei$^{12}$, F.~Weidner$^{60}$, S.~P.~Wen$^{1}$, D.~J.~White$^{59}$, U.~Wiedner$^{4}$
% ...
    \\
    \vspace{0.2cm}
    (BESIII Collaboration)\\
    \vspace{0.2cm} {\it
        $^{1}$ Institute of High Energy Physics, Beijing 100049, People's Republic of China\\
% ...
$^{4}$ Bochum Ruhr-University, D-44780 Bochum, Germany\\
% ...
$^{10}$ G.I. Budker Institute of Nuclear Physics SB RAS (BINP), Novosibirsk 630090, Russia\\
% ...
$^{12}$ Guangxi Normal University, Guilin 541004, People's Republic of China\\
% ...
$^{28}$ Johannes Gutenberg University of Mainz, Johann-Joachim-Becher-Weg 45, D-55099 Mainz, Germany\\
% ...
$^{32}$ Liaoning Normal University, Dalian 116029, People's Republic of China\\
% ...
$^{49}$ Southeast University, Nanjing 211100, People's Republic of China\\
        $^{50}$ State Key Laboratory of Particle Detection and Electronics, Beijing 100049, Hefei 230026, People's Republic of China\\
        $^{51}$ Sun Yat-Sen University, Guangzhou 510275, People's Republic of China\\
% ...
$^{53}$ Tsinghua University, Beijing 100084, People's Republic of China\\
% ...
$^{55}$ University of Chinese Academy of Sciences, Beijing 100049, People's Republic of China\\
% ...
$^{58}$ University of Jinan, Jinan 250022, People's Republic of China\\
        $^{59}$ University of Manchester, Oxford Road, Manchester, M13 9PL, United Kingdom\\
        $^{60}$ University of Muenster, Wilhelm-Klemm-Str. 9, 48149 Muenster, Germany\\
% ...
$^{63}$ University of Science and Technology of China, Hefei 230026, People's Republic of China\\
% ...
$^{66}$ University of Turin and INFN, (A)University of Turin, I-10125, Turin, Italy; (B)University of Eastern Piedmont, I-15121, Alessandria, Italy; (C)INFN, I-10125, Turin, Italy\\
        $^{67}$ Uppsala University, Box 516, SE-75120 Uppsala, Sweden\\
% ...
$^{b}$ Also at the Novosibirsk State University, Novosibirsk, 630090, Russia\\
% ...
    }
}

\begin{abstract}
    {
        The cross section of the process $\processopp$ is measured at nineteen center-of-mass energies from 2.0 to 3.08~GeV using data collected with the BESIII detector at the BEPCII storage ring. A resonant structure around 2.20~GeV is observed with statistical significance larger than 5$\sigma$. Using a coherent fit to the cross section line shape, the mass and width are determined to be \resultoppcodemass\ and \resultoppcodewidth, respectively, where the first uncertainties are statistical and the second ones are systematic.
    }
\end{abstract}
\maketitle

\section{Introduction}
{
The process $\ee\ra V\pi\pi$, where $V$ denotes a vector meson state, has been widely studied and provides an important arena for the measurements of resonant structures. For example, there are bottomonium states in the process $\ee\ra\Upsilon(nS)\pip\pim$~\cite{3_1_ypipi}, charmonium states in the $\ee\ra J/\psi\pi\pi$ and $\psi(2S)\pi\pi$ processes~\cite{3_1_jpsipipi_1,3_1_jpsipipi_2,3_1_psippipi_1,3_1_psippipi_2}, and $\phi(2170)$ in $\ee\ra \phi\pip\pim$~\cite{3_1_phipipi_belle,3_1_phipipi_bes}. Hence, it is natural to search for vector mesons in $\ee\ra\omega\pi\pi$ processes. According to isospin conservation in strong interactions and the Clebsch-Gordon coefficients involved, of the two intermediate-state isospin possibilities, $I = 0$ and $I = 1$, both are allowed for the process $\processopm$, while only $I = 0$ is allowed for $\ee\ra\omega\pi^{0}\pi^{0}$, which makes the $\omega\piz\piz$ channel most suitable to search for an intermediate isoscalar resonance.

The \babar\ Collaboration has used the initial state radiation method to measure the cross sections of various processes in the low energy region below 2.2~GeV~\cite{3_1_babar_opm,3_1_babar_opp}.  More recently they have expanded their measurements up to 2.5~GeV and used the results in conjunction with their previous results to investigate the nature of the resonance observed by the BESIII Collaboration in the $e^+ e^- \to K^+ K^-$ cross section near 2.2~GeV~\cite{3_3_kk}. Among the cross sections measured by \babar\ were the processes $\processopm$ and $\omega\pi^{0}\pi^{0}$~\cite{3_1_babar_opm,3_1_babar_opp}, where they reported a resonant structure with a mass of $2265\pm20\unitmmev$, a width of $75^{+125}_{-27}\unitemev$, and a significance of $2.6\sigma$, by combining the $\opp$ and $\opm$ channels~\cite{3_1_babar_summary}.

Since $\phi\ra\opp$ is an Okubo-Zweig-Iizuka (OZI) suppressed process, a resonant structure in $\processopp$ is more likely to be an $\omega$ excited state than a $\phi$ excited state. According to the Particle Data Group (PDG)~\cite{1_PDG}, there are three $\omega$ excited state candidates around 2.2~GeV, $\omega(2205)$~\cite{2_omega_2205}, $\omega(2290)$~\cite{2_omega_2290} and $\omega(2330)$~\cite{2_omega_2330}, which are not fully understood yet. Reference~\cite{2_omega_theory} predicts these to be $n^{3}S_{1}$ states for $\omega(2290)$ and $\omega(2330)$ and an $n^{3}D_{1}$ state for $\omega(2205)$. Further experimental investigations
are needed to disentangle this scenario.

In this paper, the Born cross sections of the process $\processopp$ are measured with data samples collected at nineteen center-of-mass energies ($\sqrt{s}$) from 2.0 to 3.08~GeV corresponding to a total integrated luminosity of $\luminosity$.  With the same data samples, several other hadronic processes have been used to search for excited meson states above 2.0~GeV, including $\ee\ra\eta'\pip\pim$, $\omega\piz$, $\omega\eta$, etc~\cite{3_3_etappipi, 3_3_kk, 3_3_kkpipi, 3_3_omegaetaomegapi, 3_3_phieta, 3_3_phietap, 3_3_phikk}.
}

\section{Detector and data sample}
{
The BESIII detector~\cite{4_1_detector_1} records symmetric $e^+e^-$ collisions provided by the BEPCII storage ring~\cite{4_1_detector_2}, which operates in the center-of-mass energy range from 2.0 to 4.95~GeV. BESIII has collected large data samples in this energy region~\cite{4_1_detector_3}. The cylindrical core of the BESIII detector covers 93\% of the full solid angle and consists of a helium-based multilayer drift chamber~(MDC), a plastic scintillator time-of-flight system~(TOF), and a CsI(Tl) electromagnetic calorimeter~(EMC), which are all enclosed in a superconducting solenoidal magnet providing a 1.0~T magnetic field. The solenoid is supported by an octagonal flux-return yoke with resistive plate counter muon identification modules interleaved with steel. The charged-particle momentum resolution at $1~{\rm GeV}/c$ is $0.5\%$, and the $dE/dx$ resolution is $6\%$ for electrons from Bhabha scattering. The EMC measures photon energies with a resolution of $2.5\%$ ($5\%$) at $1$~GeV in the barrel (end cap) region. The time resolution in the TOF barrel region is 68~ps, while that in the end cap region is 110~ps.

Simulated data samples produced with a {\sc geant4}-based~\cite{4_2_gen_geant4} Monte Carlo (MC), which includes the geometric description of the BESIII detector and the detector response, are used to determine detection efficiencies and to estimate backgrounds. The known decay modes are modeled with {\sc evtgen}~\cite{4_2_gen_evtgen} using branching fractions taken from the PDG~\cite{1_PDG}. Final state radiation~(FSR) from charged final state particles is incorporated using {\sc photos}~\cite{4_2_gen_photons}, and initial state radiation (ISR) is incorporated using ConExc~\cite{4_2_gen_conexc}. The $\opp$ state is simulated by using a uniformly distributed phase space (PHSP) model. The decay of $\omega$ to $\pip\pim\piz$ is simulated by using a Dalitz plot analysis as described in Ref.~\cite{4_2_gen_omegadalitz}. Inclusive MC events for studying background contamination are generated using a hybrid generator~\cite{4_2_gen_hybrid}, which includes hadronic events and background events.
}

\section{Event selection and background analysis}
{
\label{event_selection}
For the process $\processopp$, with subsequent decays $\omega\rightarrow\pipipi$ and $\piz\rightarrow\gamma\gamma$, candidate events are required to have two reconstructed charged tracks and at least six reconstructed photons. Charged tracks detected in the MDC are required to be within a polar angle ($\theta$) range of $|\rm{cos\theta}|<0.93$, where $\theta$ is defined with respect to the $z$-axis, which is the symmetry axis of the MDC. The distance of closest approach to the interaction point must be less than 10\,cm along the $z$-axis and less than 1\,cm in the transverse plane. Photon candidates are identified using showers in the EMC. The deposited energy of each shower must be more than 25~MeV in the barrel region ($|\!\cos \theta|< 0.80$) and more than 50~MeV in the end cap region ($0.86 <|\!\cos \theta|< 0.92$).
To exclude showers that originate from charged tracks, the angle between the line joining the interaction point (IP) to the position of the selected shower and a line joining the IP to the point where any charged track is projected to intersect the EMC must be greater than 10 deg. To suppress electronic noise and showers unrelated to the event, the difference between the EMC time and the event start time is required to be within [0, 700]\,ns.

Particle identification~(PID) for charged tracks combines measurements of $dE/dx$ in the MDC and the flight time in the TOF to form likelihoods $\mathcal{L}(h)~(h=p,K,\pi)$ for each hadron $h$ hypothesis. Tracks are identified as pions when the pion hypothesis has the greatest likelihood [$\mathcal{L}(\pi)>\mathcal{L}(K)$ and $\mathcal{L}(\pi)>\mathcal{L}(p)$]. Two identified oppositely charged pions are required and then used in a vertex fit. Only events with two oppositely charged pions satisfying the vertex fit are selected.

To suppress background events, a four-constraint (4C) kinematic fit imposing four-momentum conservation is performed under the hypothesis $\ee\ra\opp\ra\pip\pim\gamma\gamma\gamma\gamma\gamma\gamma$, with $\chisqfour<100$ required, where $\chisqfour$ is the $\chisq$ from the kinematic fit. For events with more than six photon candidates, the combination of six different photons with the smallest $\chisqfour$ is retained and $\chisqfour$ should be less than 100 as well. Three photon pairs corresponding to the three $\piz$ candidates are selected by choosing the combination with the smallest value of
$\chi^{2}_{\pi^0\pi^0\pi^0} =
    (M_{\gamma_1\gamma_2}-M_{\pi ^0}^{\rm PDG})^2/\sigma_{\gamma_1\gamma_2}^{2}
    +(M_{\gamma_3\gamma_4}-M_{\pi^0}^{\rm PDG})^2/\sigma_{\gamma_3\gamma_4}^{2}
    +(M_{\gamma_5\gamma_6}-M_{\pi^0}^{\rm PDG})^2/\sigma_{\gamma_5\gamma_6}^{2}$,
where $M^{\rm PDG}_{\pi^0}$ is the mass of $\piz$ from the PDG~\cite{1_PDG}, while $M_{\gamma_{i}\gamma_{j}}$ and $\sigma_{\gamma_{i}\gamma_{j}}$ are the invariant mass of $\gamma_{i}\gamma_{j}$ and its calculated standard deviations from MC samples. Of the three $\piz$ mesons, the one with the minimum $|M_{\pipipi}-M^{\rm PDG}_{\omega}|$ is assigned to be from the $\omega$ decay and tagged as $\pizone$, where $M^{\rm PDG}_{\omega}$ is the mass of $\omega$ from the PDG~\cite{1_PDG}. The two photons used to reconstruct $\pizone$ are tagged as $\gamma_{1}$ and $\gamma_{2}$. The other two $\piz$ mesons are tagged as $\piztwo$ and $\pizthree$ according to $M_{\omega\piztwo}<M_{\omega\pizthree}$, where $M_{\omega\piztwo}$ and $M_{\omega\pizthree}$ represent the invariant mass of $\omega\piztwo$ and $\omega\pizthree$, respectively. The photons used to reconstruct them are tagged as $\gamma_{3}$, $\gamma_{4}$, $\gamma_{5}$ and $\gamma_{6}$, respectively.

The difference between the invariant mass of the reconstructed $\piz$ and $M_{\pi ^0}^{\rm PDG}$ is required to be less than 3 times the left (right) side standard deviation: $M_{\gamma_{i}\gamma_{j}}\in[M^{\rm PDG}_{\piz} - 3\cdot\sigma_{({\rm left})\gamma_{i}\gamma_{j}},\ M^{\rm PDG}_{\piz} + 3 \cdot\sigma_{({\rm right})\gamma_{i}\gamma_{j}}]$, where $\sigma_{({\rm left})}$ and $\sigma_{({\rm right})}$ are the quadratic means of the difference of the mass of reconstructed $\piz$ and $M_{\pi^0}^{\rm PDG}$ for $\piz$ candidates with mass above and below $M_{\pi ^0}^{\rm PDG}$, respectively.

The invariant mass distribution of the reconstructed $\omega$ candidates at $\energy[2.1250]$ is shown in Fig.~\ref{figure:1:omega}. There are contributions from both PHSP MC and inclusive backgrounds. In this distribution, $|M_{\pipipionefull}-M^{\rm PDG}_{\omega}| < 0.05\unitmgev$ is chosen as the signal region, as shown in Fig.~\ref{figure:1:omega}, while $|M_{\pipipionefull}-M^{\rm PDG}_{\omega}|\in [0.10,0.20]\unitmgev$ is chosen as the sideband region to estimate backgrounds.

\begin{figure}[!hbpt]
    \begin{center}
        \includegraphics[width=0.45\textwidth]{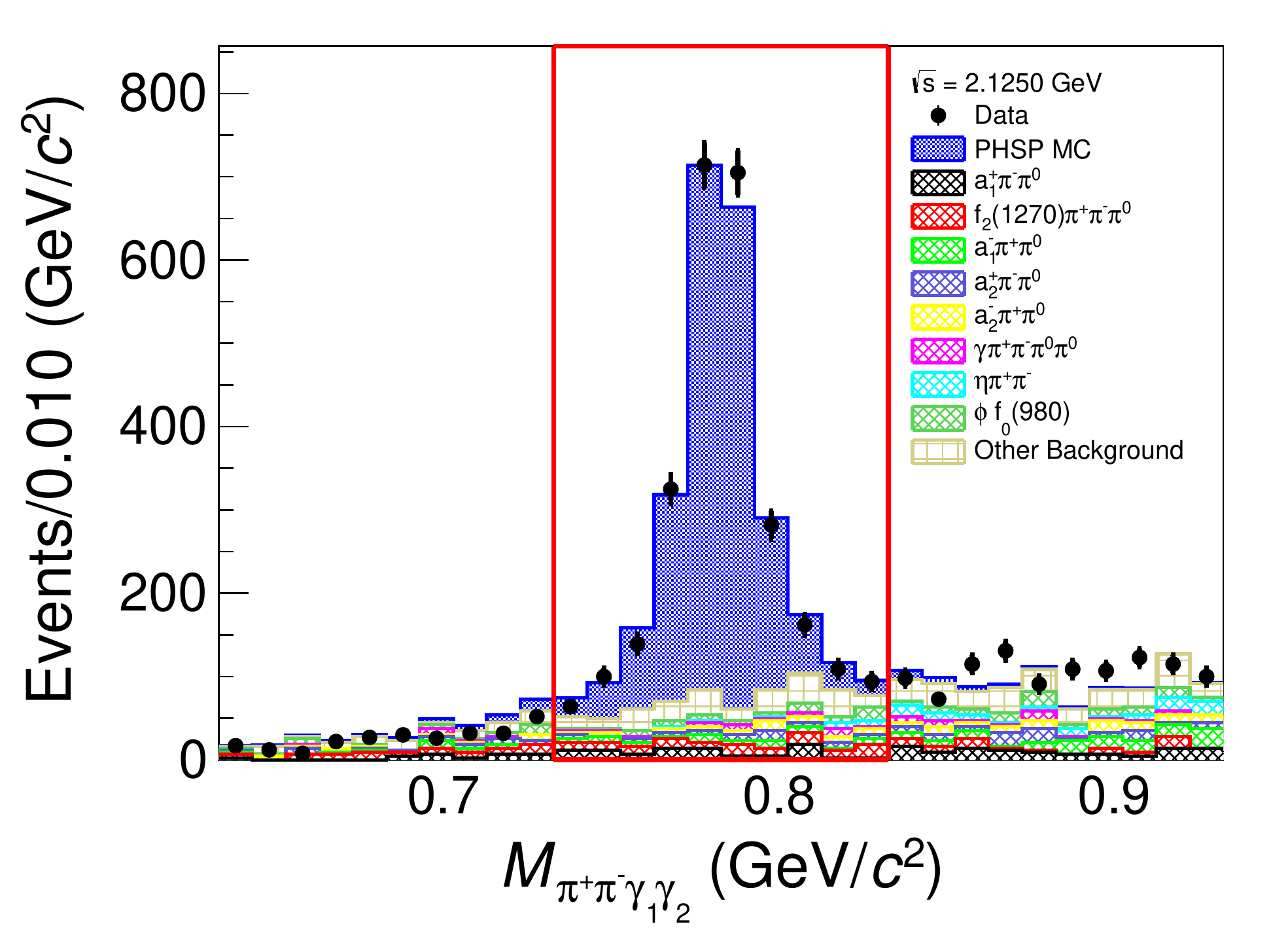}
    \end{center}
    \caption{The invariant mass distribution of $\pipipionefull$ at $\energy$. The black dots with error bars are data. The blue histogram represents the contribution of PHSP MC. Other colored histograms represent contributions of backgrounds from inclusive MC. The red box indicates the signal region. Backgrounds are normalized according to the estimated cross section of each process.}
    \label{figure:1:omega}
\end{figure}

Inclusive MC events are selected with the same event selection criteria. Detailed event type analysis over these events with TopoAna~\cite{4_2_gen_topoana} shows that the dominant backgrounds come from processes with $\pip\pim\piz\piz\piz$ final states but through different intermediate states. However, no peaking background appears under the $\omega$ resonance.
}

\section{Born cross section measurement}
{
The Born cross section of $\processopp$ is calculated from

\begin{equation}
    \begin{aligned}
        \sigma^{\rm B} = \frac{N_{\rm signal}}{\mathcal{L} \cdot \varepsilon\cdot \mathcal{B}_{\omega\ra\pipipi}\cdot \mathcal{B}^{3}_{\piz\ra\gamma\gamma}\cdot (1+\delta)},
    \end{aligned}
\end{equation}
where $\mathcal{L}$ is the luminosity, $N_{\rm signal}$ is the signal yield, $\varepsilon$ is the detection efficiency and $\mathcal{B}_{\omega\ra\pipipi}$ and $\mathcal{B}_{\piz\ra\gamma\gamma}$ are branching fractions taken from the PDG~\cite{1_PDG}. The product of the ISR correction factor times the vacuum-polarization (VP) correction factor is represented by $1+\delta$.

\begin{figure*}[!htbp]
    \begin{center}
        \begin{overpic}[width=0.31\textwidth]{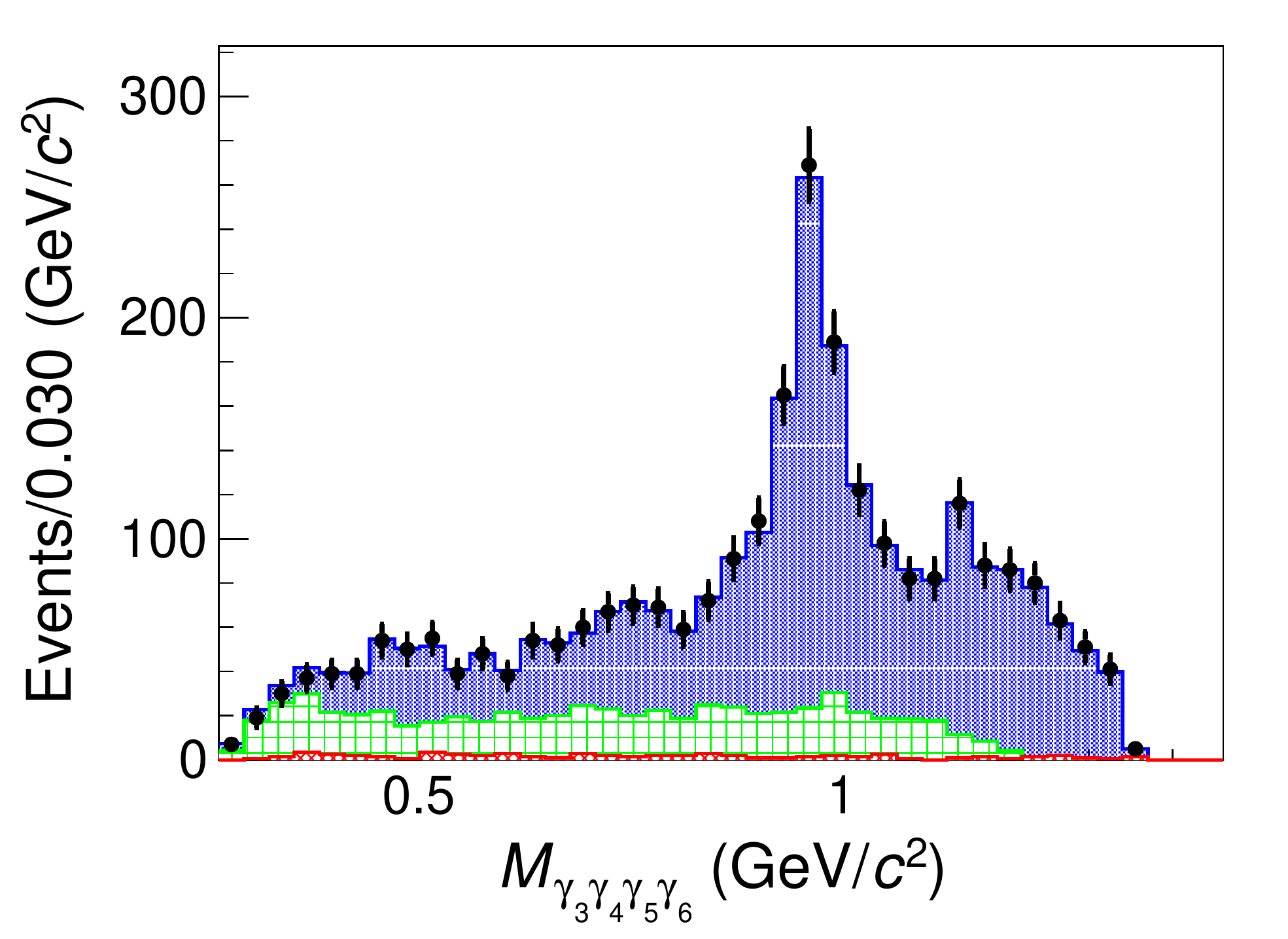}
            \put(25,60){\small\bfseries{(a)}}
        \end{overpic}
        \begin{overpic}[width=0.31\textwidth]{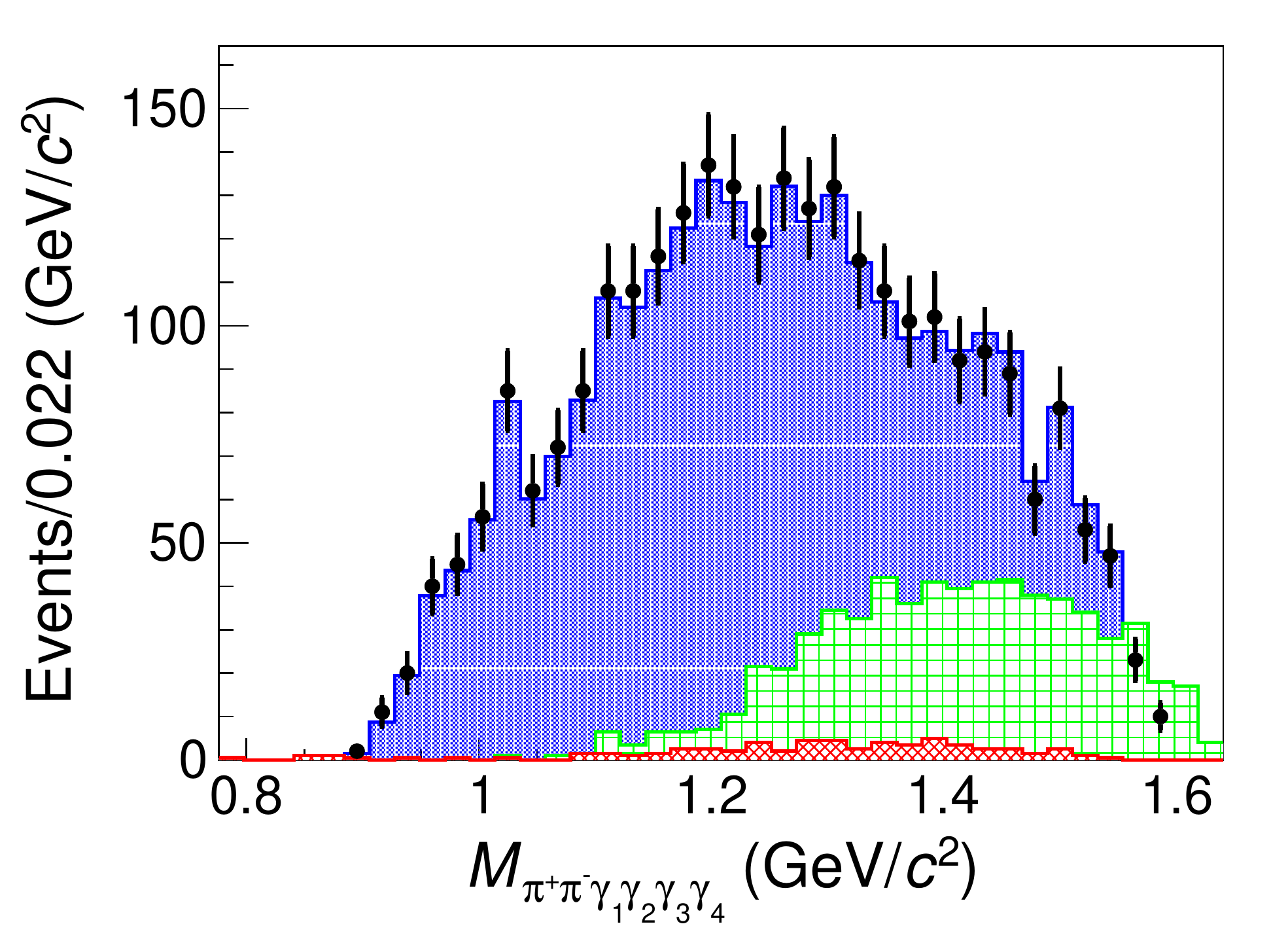}
            \put(25,60){\small\bfseries{(b)}}
        \end{overpic}
        \begin{overpic}[width=0.31\textwidth]{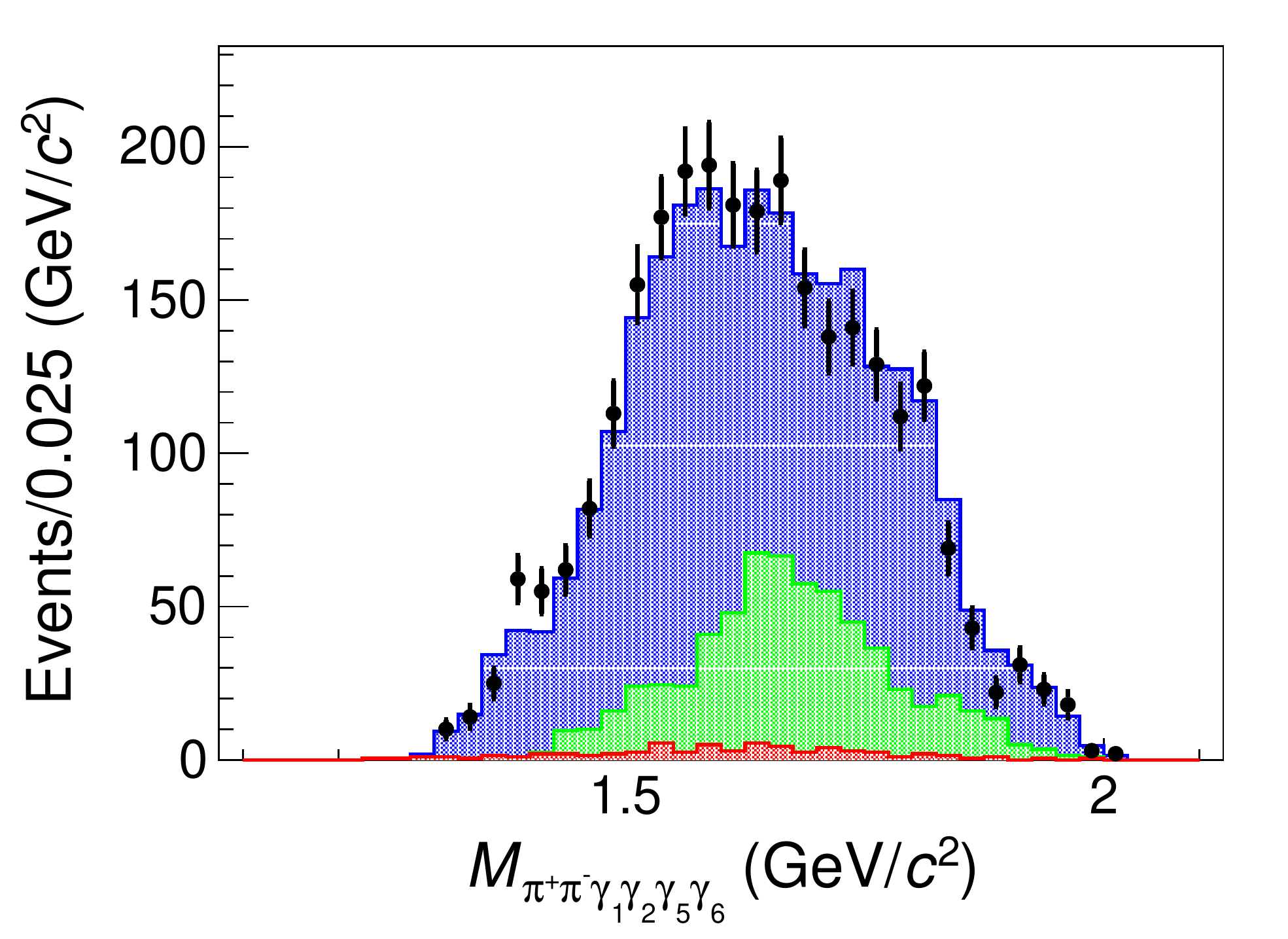}
            \put(25,60){\small\bfseries{(c)}}
        \end{overpic}
        \begin{overpic}[width=0.31\textwidth]{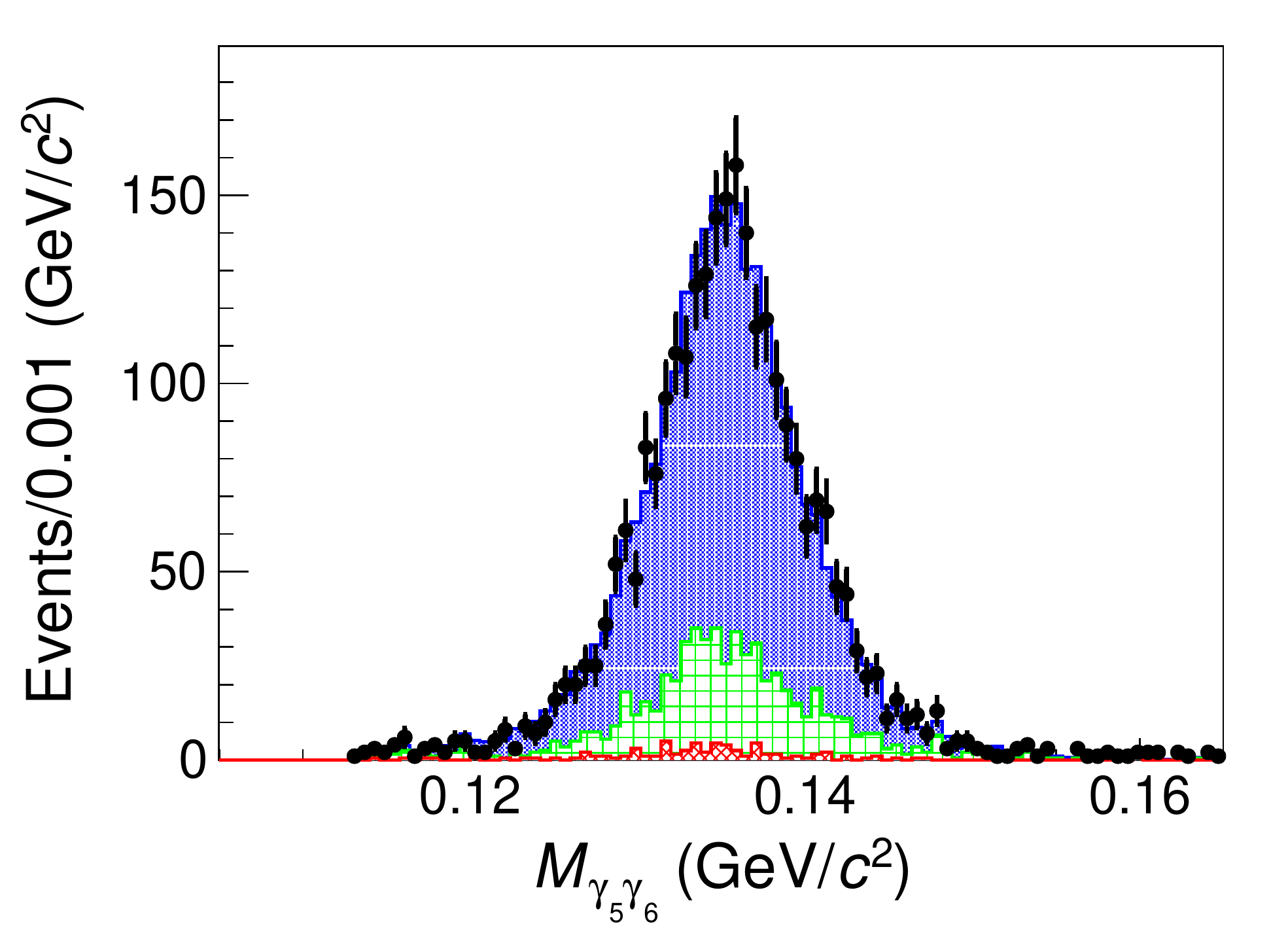}
            \put(25,60){\small\bfseries{(d)}}
        \end{overpic}
        \begin{overpic}[width=0.31\textwidth]{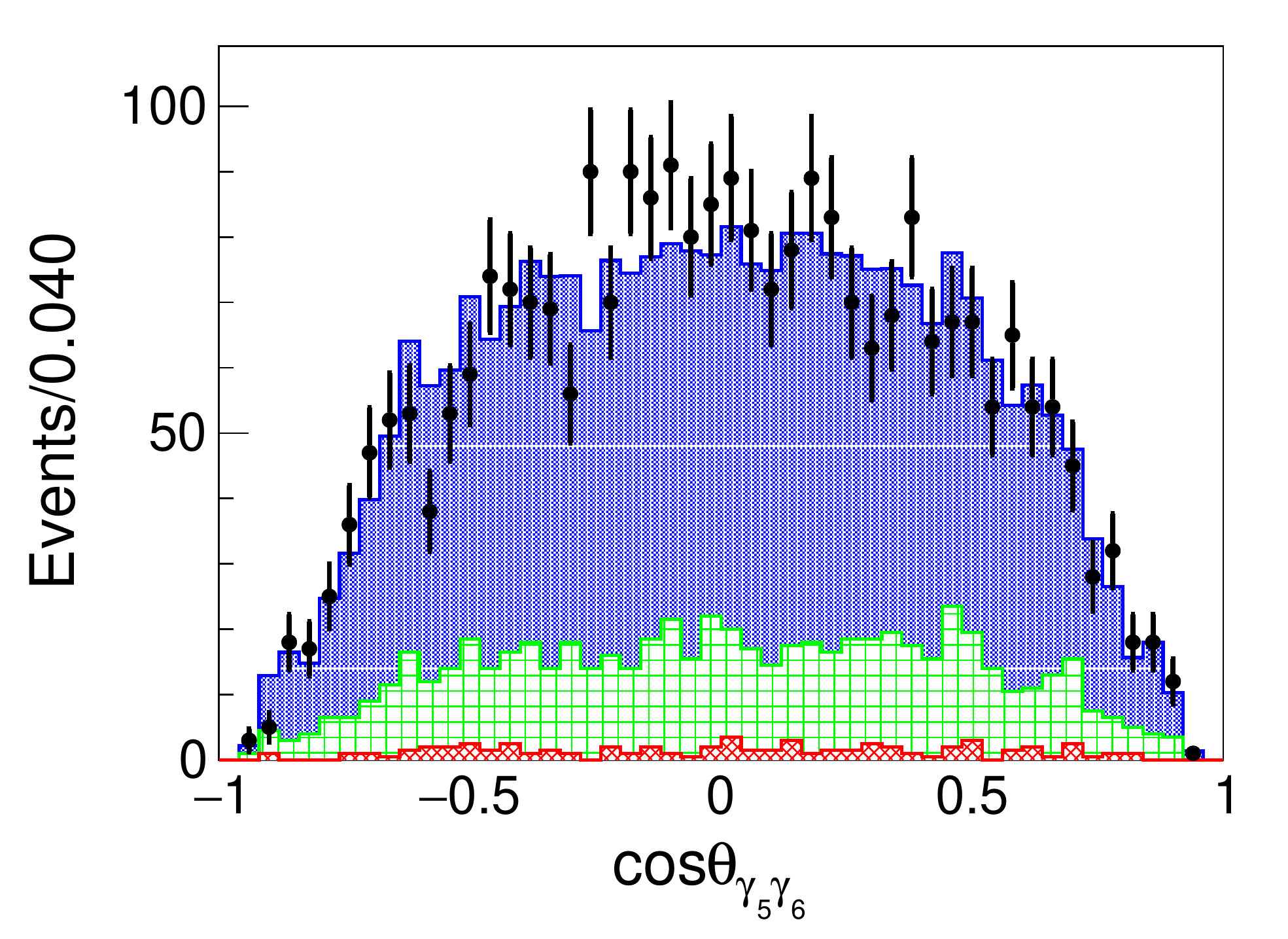}
            \put(25,60){\small\bfseries{(e)}}
        \end{overpic}
        \caption{The invariant mass distributions of (a)
            $\piztwofull\pizthreefull$, (b) $\pipipionefull\piztwofull$,
            (c) $\pipipionefull\pizthreefull$, (d) $\pizthreefull$, and
            (e) ${\rm cos}\theta_{\pizthreefull}$ at $\energy$, where
            $\theta_{\pizthreefull}$ is the polar angle of the
            $\pizthreefull$ system defined with respect to the $z$-axis.
            In both (d) and (e), the distribution of $\pizthreefull$ is
            shown as an example of the three possible $\gamma\gamma$
            combinations.  The black dots with error bars represent
            data. The blue histogram represents the contribution of the
            weighted signal MC. The red and green histograms represent
            the contributions of estimated backgrounds, which are given
            by one half of the sum of the left and right sidebands in
            the $\pi^+ \pi^- \gamma_1 \gamma_2$ invariant mass
            distribution, respectively. Histograms are stacked above one
            another to compare with data. The histograms are normalized
            according to the number of events in the histograms.}
        \label{figure:2:weight}
    \end{center}
\end{figure*}

The PHSP MC samples are found to strongly deviate from the data. To obtain a more reliable detection efficiency, the PHSP MC events are weighted according to the multidimensional distribution. Two-body invariant mass distributions and pion angular distributions are suitable to correct the impact of intermediate processes. Since $M_{\pipipionefull\piztwofull}$ and $M_{\pipipionefull\pizthreefull}$ are strongly correlated (the relationship chosen in Sec.~\ref{event_selection} is not sufficient to separate the role of $\piztwo$ and $\pizthree$), two-dimensional distribution of $M_{\pipipionefull\piztwofull}$ versus $M_{\piztwofull\pizthreefull}$ are found to be better to satisfy the consistency of data and PHSP MC. The weight factor is the ratio between data and the PHSP MC in this distribution with 40 bins for each dimension. It is defined as $w=(n_{\rm Data}-n_{\rm Sideband})/n_{\rm MC}$, where $n$ denotes the number of events in the corresponding bin. Good agreement between data and weighted MC distributions is observed, as shown in Fig.~\ref{figure:2:weight}. The detection efficiency ($\varepsilon$) is taken as the total weight of selected events divided by the total weight of generated events.

The signal yield of $\processopp$ is obtained by fitting the
$\pipipionefull$ mass spectrum with an unbinned maximum likelihood
method. The contribution of background events is described by a
second order polynomial function, and the $\omega$ signal is described by the MC-simulated
shape convolved with a Gaussian function which accounts for the
difference between MC and data. Figure \ref{figure:3:yield} shows the
fitted $\pipipionefull$ mass spectrum for the data sample at
$\energy[2.1250]$.

\begin{figure}[!hbpt]
    \begin{center}
        \includegraphics[width=0.45\textwidth]{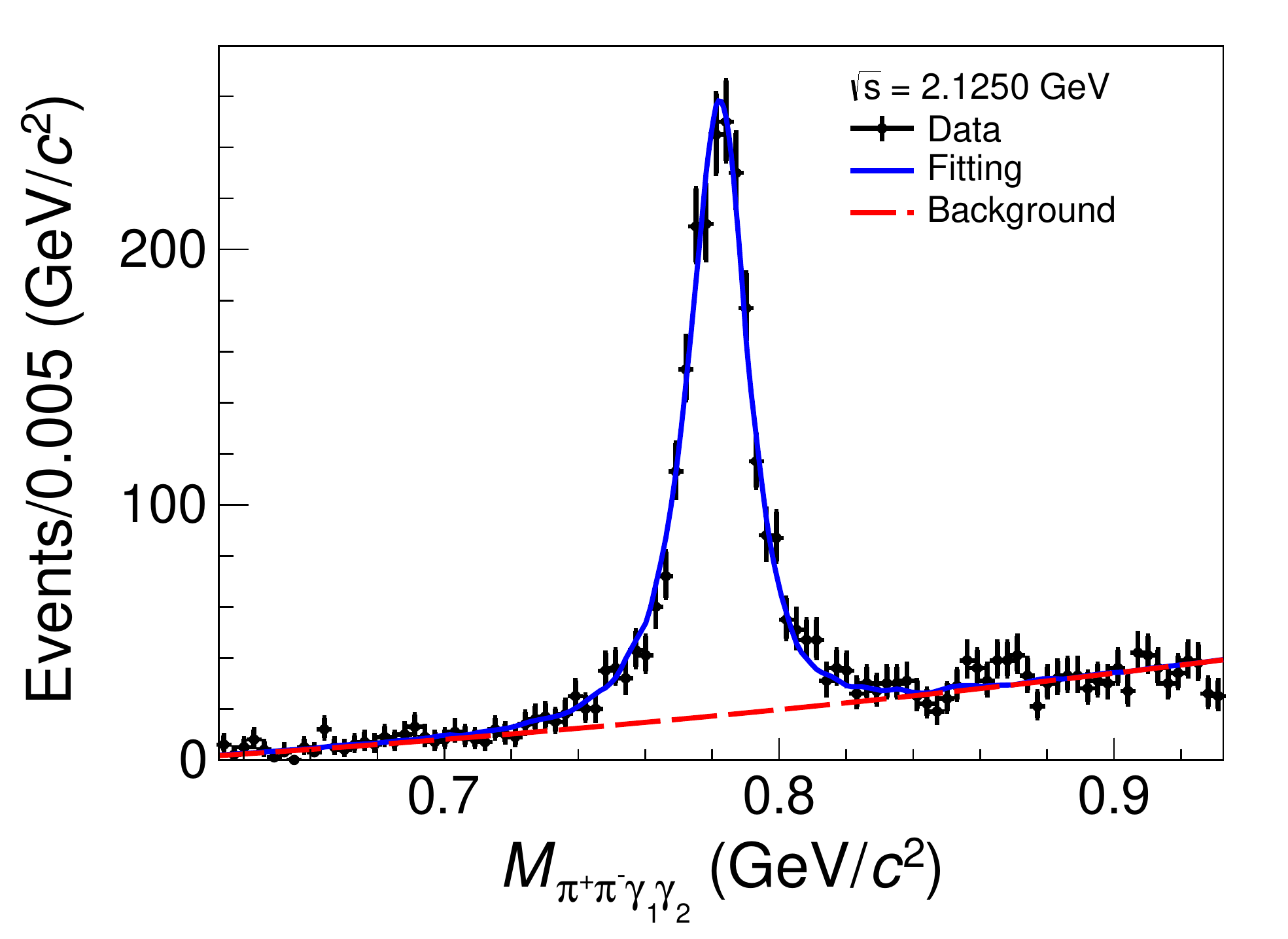}
    \end{center}
    \caption{Fitting result for the $\pipipionefull$ mass
        spectrum for the data sample at $\energy[2.1250]$. Dots with error bars represent data. The blue solid line
        represents the fitting function, and the red dashed line represents the background.}
    \label{figure:3:yield}
\end{figure}

\begin{table*}[!htbp]
    \begin{center}
        \caption{Born cross sections for $\processopp$. The columns
            represent center-of-mass energy ($\sqrt{s}$), signal yield,
            luminosity, detection efficiency, ISR and VP corrections and
            calculated Born cross section. The first uncertainties for
            the Born cross section are statistical, and the second ones are
            systematic. The uncertainties for the signal yield are
            statistical only.}
        \label{table_borncrosssection}
        \begin{small}
            \begin{tabular}{C{2cm}R{0.8cm}PR{0.4cm}R{1cm}@{.}L{0.8cm}C{1.5cm}C{1.5cm}R{1cm}PR{0.6cm}PR{0.6cm}r}
                \hline\hline
                $\sqrt{s}$ (GeV) & \multicolumn{2}{c}{$N_{\rm signal}$} & \multicolumn{2}{c}{$\mathcal{L}\ (pb^{-1})$} & $\varepsilon$ & $1+\delta$ & \multicolumn{4}{c}{Born cross section (pb)}                                \\
                \hline
                2.0000           & 273                                  & 18                                           & 10            & 1          & 0.12                                        & 0.89 & 296.9 & 20.1 & 20.2 & \\
                2.0500           & 86                                   & 10                                           & 3             & 34         & 0.12                                        & 0.94 & 276.3 & 32.7 & 18.9 & \\
                2.1000           & 249                                  & 18                                           & 12            & 2          & 0.11                                        & 0.97 & 218.1 & 15.9 & 15.2 & \\
                2.1250           & 2144                                 & 53                                           & 108           & 0          & 0.11                                        & 0.98 & 211.2 & 5.2  & 14.4 & \\
                2.1500           & 54                                   & 8                                            & 2             & 84         & 0.11                                        & 0.98 & 206.6 & 31.6 & 14.1 & \\
                2.1750           & 242                                  & 18                                           & 10            & 6          & 0.11                                        & 0.98 & 227.9 & 16.6 & 15.5 & \\
                2.2000           & 308                                  & 20                                           & 13            & 7          & 0.12                                        & 0.97 & 229.9 & 14.6 & 15.6 & \\
                2.2324           & 224                                  & 17                                           & 11            & 9          & 0.11                                        & 1.04 & 185.6 & 13.7 & 12.7 & \\
                2.3094           & 207                                  & 17                                           & 21            & 1          & 0.10                                        & 1.13 & 102.1 & 8.2  & 7.0  & \\
                2.3864           & 194                                  & 19                                           & 22            & 5          & 0.10                                        & 1.14 & 88.2  & 8.5  & 6.0  & \\
                2.3960           & 526                                  & 27                                           & 66            & 9          & 0.10                                        & 1.14 & 76.7  & 3.9  & 5.2  & \\
                2.6444           & 207                                  & 16                                           & 33            & 7          & 0.10                                        & 1.25 & 59.4  & 4.6  & 4.0  & \\
                2.6464           & 188                                  & 16                                           & 34            & 0          & 0.10                                        & 1.28 & 54.8  & 4.5  & 3.7  & \\
                2.9000           & 296                                  & 19                                           & 105           & 0          & 0.10                                        & 1.34 & 25.4  & 1.6  & 1.7  & \\
                2.9500           & 32                                   & 9                                            & 15            & 9          & 0.10                                        & 1.33 & 19.8  & 5.5  & 1.4  & \\
                2.9810           & 52                                   & 8                                            & 16            & 1          & 0.10                                        & 1.38 & 30.4  & 4.7  & 2.1  & \\
                3.0000           & 28                                   & 6                                            & 15            & 9          & 0.10                                        & 1.29 & 15.5  & 3.6  & 1.1  & \\
                3.0200           & 30                                   & 6                                            & 17            & 3          & 0.09                                        & 1.34 & 15.6  & 3.1  & 1.1  & \\
                3.0800           & 156                                  & 14                                           & 126           & 0          & 0.09                                        & 1.34 & 12.7  & 1.2  & 0.9  & \\
                \hline\hline
            \end{tabular}
        \end{small}
    \end{center}
\end{table*}

The ISR and VP effects are incorporated by
ConExc~\cite{4_2_gen_conexc}, which provides ISR and VP factors
depending on the input cross section. An iterative procedure is performed, with
comparison between the input cross sections and the measured ones, until the
difference of $(1+\delta)$ is less than 1\% between the last two
iterations.

The Born cross sections for all nineteen energies together with all values used in the measurement are shown in Table~\ref{table_borncrosssection}.
}

\section{Systematic uncertainty}
 {
  Several sources of systematic uncertainties, which include the
  luminosity measurements, tracking efficiency, PID efficiency,
  photon detection, kinematic fit, $\gamma\gamma$ mass requirement,
  fitting procedure, branching fractions of intermediate state decays
  and the ISR and VP corrections, are considered in this analysis.

  \begin{table*}[!htbp]
      \begin{center}
          \caption{Systematic uncertainties (in percentage) in this analysis. The columns represent center-of-mass energy ($\sqrt{s}$), the uncertainty associated with luminosity, charged track selection, PID, photon selection, branching fraction, 4C kinematic fit, reweight procedure, background shape, signal shape, $\gamma\gamma$ invariant mass window, $\pipipionefull$ invariant mass window and $1+\delta$ calculation. The last column is the total systematic uncertainty.}
          \label{table_systematicuncertainty}
          \begin{small}
              \begin{tabular}{cccccccccccccc}
                  \hline\hline
                                   &               & Charged track &     & Photon    & Branching & 4C            & Reweight  & Background & Signal & $M_{\gamma\gamma}$ & $M_{\pipipionefull}$ &            &       \\
                  $\sqrt{s}$ (GeV) & $\mathcal{L}$ & selection     & PID & selection & fraction  & kinematic fit & procedure & shape      & shape  & window             & window               & $1+\delta$ & Total \\
                  \hline
                  2.0000           & 1.0           & 2.0           & 2.0 & 6.0       & 0.7       & 0.57          & 0.35      & 0.22       & 0.13   & 0.18               & 0.18                 & 0.5        & 6.81  \\
                  2.0500           & 1.0           & 2.0           & 2.0 & 6.0       & 0.7       & 0.55          & 0.66      & 0.10       & 0.18   & 0.16               & 0.18                 & 0.5        & 6.83  \\
                  2.1000           & 1.0           & 2.0           & 2.0 & 6.0       & 0.7       & 0.66          & 0.33      & 0.17       & 1.43   & 0.11               & 0.11                 & 0.5        & 6.96  \\
                  2.1250           & 1.0           & 2.0           & 2.0 & 6.0       & 0.7       & 0.36          & 0.14      & 0.19       & 0.41   & 0.26               & 0.17                 & 0.5        & 6.80  \\
                  2.1500           & 1.0           & 2.0           & 2.0 & 6.0       & 0.7       & 0.68          & 0.66      & 0.12       & 0.45   & 0.14               & 0.14                 & 0.5        & 6.85  \\
                  2.1750           & 1.0           & 2.0           & 2.0 & 6.0       & 0.7       & 0.46          & 0.45      & 0.11       & 0.18   & 0.11               & 0.18                 & 0.5        & 6.80  \\
                  2.2000           & 1.0           & 2.0           & 2.0 & 6.0       & 0.7       & 0.52          & 0.46      & 0.19       & 0.14   & 0.15               & 0.16                 & 0.5        & 6.81  \\
                  2.2324           & 1.0           & 2.0           & 2.0 & 6.0       & 0.7       & 0.47          & 0.48      & 0.14       & 0.15   & 0.17               & 0.52                 & 0.5        & 6.82  \\
                  2.3094           & 1.0           & 2.0           & 2.0 & 6.0       & 0.7       & 0.36          & 0.46      & 0.93       & 0.16   & 0.18               & 0.16                 & 0.5        & 6.86  \\
                  2.3864           & 1.0           & 2.0           & 2.0 & 6.0       & 0.7       & 0.34          & 0.53      & 0.14       & 0.17   & 0.11               & 0.13                 & 0.5        & 6.80  \\
                  2.3960           & 1.0           & 2.0           & 2.0 & 6.0       & 0.7       & 0.54          & 0.34      & 0.11       & 0.24   & 0.47               & 0.10                 & 0.5        & 6.82  \\
                  2.6444           & 1.0           & 2.0           & 2.0 & 6.0       & 0.7       & 0.39          & 0.32      & 0.17       & 0.26   & 0.32               & 0.36                 & 0.5        & 6.81  \\
                  2.6464           & 1.0           & 2.0           & 2.0 & 6.0       & 0.7       & 0.60          & 0.73      & 0.13       & 0.29   & 0.10               & 0.16                 & 0.5        & 6.84  \\
                  2.9000           & 1.0           & 2.0           & 2.0 & 6.0       & 0.7       & 0.23          & 0.35      & 0.17       & 0.14   & 0.17               & 0.15                 & 0.5        & 6.78  \\
                  2.9500           & 1.0           & 2.0           & 2.0 & 6.0       & 0.7       & 0.49          & 0.95      & 0.18       & 1.48   & 0.61               & 0.77                 & 0.5        & 7.08  \\
                  2.9810           & 1.0           & 2.0           & 2.0 & 6.0       & 0.7       & 0.50          & 1.17      & 0.14       & 0.13   & 0.13               & 0.11                 & 0.5        & 6.89  \\
                  3.0000           & 1.0           & 2.0           & 2.0 & 6.0       & 0.7       & 0.64          & 1.04      & 0.13       & 1.06   & 0.13               & 0.66                 & 0.5        & 6.99  \\
                  3.0200           & 1.0           & 2.0           & 2.0 & 6.0       & 0.7       & 0.26          & 1.02      & 0.14       & 0.11   & 0.12               & 0.17                 & 0.5        & 6.85  \\
                  3.0800           & 1.0           & 2.0           & 2.0 & 6.0       & 0.7       & 0.35          & 0.73      & 0.11       & 0.13   & 0.15               & 0.10                 & 0.5        & 6.82  \\
                  \hline\hline
              \end{tabular}
          \end{small}
      \end{center}
  \end{table*}

  (a) The integrated luminosities of the data samples used in this analysis are measured using large angle Bhabha scattering events, and the corresponding uncertainties are estimated to be 1.0\%~\cite{4_3_error_lumin}.

  (b) The uncertainty of the tracking efficiency is investigated using
  samples of the $\ee\rightarrow K^{+} K^{-}\pip\pim$
  process~\cite{3_3_kk,4_3_error_kkpipipi}. The difference
  in tracking efficiency between data and the MC simulation is
  estimated to be 1.0\% per pion. Hence, 2.0\% is taken as the
  systematic uncertainty.

  (c) To estimate the uncertainty in the PID efficiency, the same
  samples as used to investigate the tracking efficiency are
  studied. The average difference in the PID efficiency between data and
  the MC simulation is found to be 1.0\% per charged pion. Therefore,
  2.0\% is taken as the systematic uncertainty.

  (d) The uncertainty associated with the photon selection efficiency
  is studied with samples of $\ee\rightarrow K^{+}
      K^{-}\pip\pim\piz$~\cite{4_3_error_kkpipipi}. The samples
  cover the same angle and momentum ranges as in this analysis. The
  result shows that the difference in detection efficiency between
  data and MC simulation is 1.0\% per photon. The systematic
  uncertainty of six photons is fully correlated and result in 6.0\%
  uncertainty in total.

  (e) The uncertainty associated with the branching fractions of
  intermediate states are taken from the PDG as 0.7\%.

  (f) The uncertainty associated with reweight procedure comes from the choosing of bin size and the fluctuation of bin content of distribution of data. To estimated those impact, number of bins for each dimension are varied from 20 to 60 and the sampling of reweight factors is performed based on the error of bin content of data, with total number of data sample remains the same. These parallel samples of reweight factors are used to calculate detection efficiency again. The standard deviation of these parallel detection efficiencies in percentage is taken as the systematic uncertainty.

  (g) The uncertainty associated with the kinematic fit comes from the
  inconsistency of the track helix parameters between data and the MC
  simulation. The helix parameters for the charged tracks of MC
  samples are corrected to eliminate the inconsistency, as described
  in Ref.~\cite{4_3_error_helix}, and the agreement of the $\chisqfour$
  distributions between data and MC simulation is significantly
  improved. The differences of the selection efficiencies with and
  without the correction are taken as the systematic uncertainties.

  (h) The uncertainty associated with the background shape is
  estimated by the difference if a first order polynomial function
  is used for the background shape.

  (i) The uncertainty associated with the signal function in the signal
  determination is estimated by the difference if an alternative fit with
  a Breit-Wigner function convolved with a Gaussian function is used for
  the signal shape.

  (j) The uncertainty associated with the mass window of the
  $\gamma\gamma$ invariant mass distribution is estimated by changing
  the number of one-side standard deviations from 3 to 2.8 and
  3.2. The larger difference in the calculated cross section results
  is taken as the systematic uncertainty.

  (k) The uncertainty associated with the mass window of $\pipipionefull$ is estimated by changing the fitting range from  $M^{\rm PDG}_{\omega}\pm 0.15\unitenergyg$ to $M^{\rm PDG}_{\omega}\pm 0.14\unitenergyg$ and $M^{\rm PDG}_{\omega}\pm0.16\unitenergyg$. The larger difference in the result of the calculated cross section is taken as the systematic uncertainty.

  (l) The uncertainty associated with $1 + \delta$ is
  obtained from the accuracy of the radiation function, which is about
  0.5\%~\cite{4_3_error_isr}, and the contribution from the cross
  section line shape, which is estimated by varying the model
  parameters of the fit to the cross section. All parameters are
  randomly varied within their uncertainties, and the resulted
  parametrization of the line shape is used to recalculate $1+\delta$,
  $\epsilon$ and the corresponding cross section. This procedure is
  repeated 100 times, and the standard deviation of the resulting cross
  section is taken as the systematic uncertainty.

  All systematic uncertainties are summarized in Table~\ref{table_systematicuncertainty}. The total systematic uncertainty is obtained by adding all individual contributions in quadrature.
 }

\section{Line shape fitting to the cross section}
{
To study the possible structure around 2.20 GeV, the cross section $\sigma^{\rm B}(s)$ is fitted by the coherent sum of the possible resonant component together with a phase space component for the continuum contribution:

\begin{equation}
    \label{eq.fitting}
    \begin{aligned}
         & \sigma^{\rm B}(s) = |f_r(s)e^{i\phi _r}+f_c(s)|^2,                                                                                                                \\
         & f_r(s) = \frac{M_r}{\sqrt{s}}\frac{\sqrt{12\pi C \varGamma_{ee}^{r}Br\varGamma _r}}{s-M_{r}^{2}+iM_r\varGamma _r}\sqrt{\frac{\varPhi (\sqrt{s})}{\varPhi (M_r)}}, \\
         & f_c(s) = a\frac{\sqrt{\varPhi \left( \sqrt{s} \right)}}{(\sqrt{s})^{b}},
    \end{aligned}
\end{equation} where $f_r(s)$ represents the resonant component~\cite{4_4_lineshape_function}, in which $M_{r}$ and $\varGamma_{r}$ are the mass and width of the resonant structure near $2.20\unitegev$. Parameter $\varGamma_{ee}^{r}Br$ is the electric partial width times the branching fraction of the resonance decaying to $\omega\piz\piz$, $\phi_{r}$ is the relative phase between the resonant and nonresonant amplitude, and the continuum part $f_c(s)$ is parametrized by $a$ and $b$. All six parameters above are floated, while $\varPhi$ is the calculated three-body phase space factor and $C$ is a conversion constant which equals to $3.893\times 10^8\ {\rm pb\cdot GeV^2}$~\cite{3_1_babar_opm}. The results from the fit are shown in Fig.~\ref{figure:4:fit} and Table \ref{table:3:fit}. Two solutions are found. Solution (a) corresponds to the case of constructive interference between the resonant and continuum contributions and solution (b) corresponds to the case of destructive interference. The fitting quality $\chisq/{\rm ndf}$ is $22.6/13$, where ${\rm ndf}$ is the number of degrees of freedom.

\begin{figure}[!hbpt]
    \begin{center}
        \begin{overpic}[width=0.45\textwidth]{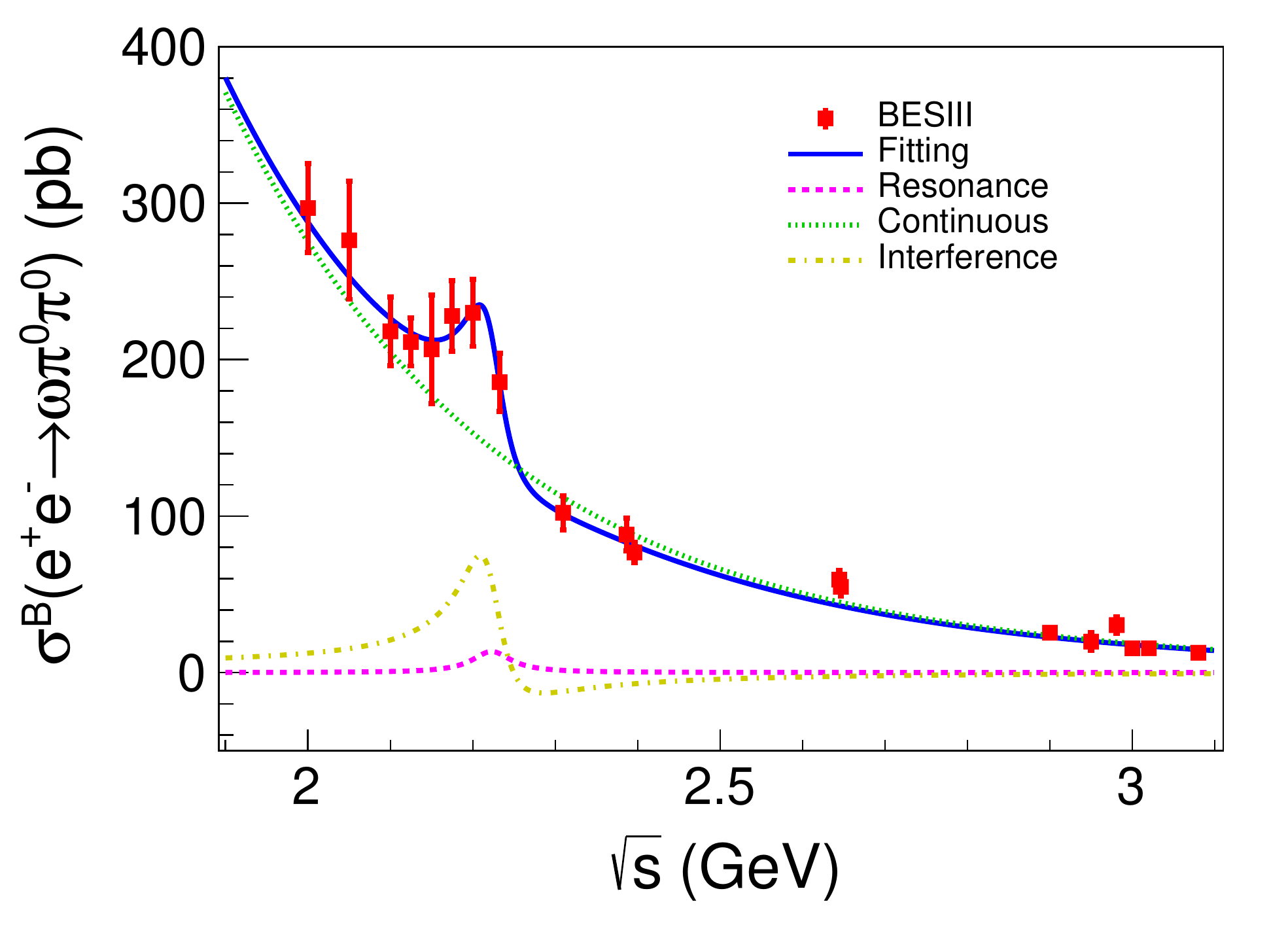}
            \put(25,60){\small\bfseries{(a)}}
        \end{overpic}
        \begin{overpic}[width=0.45\textwidth]{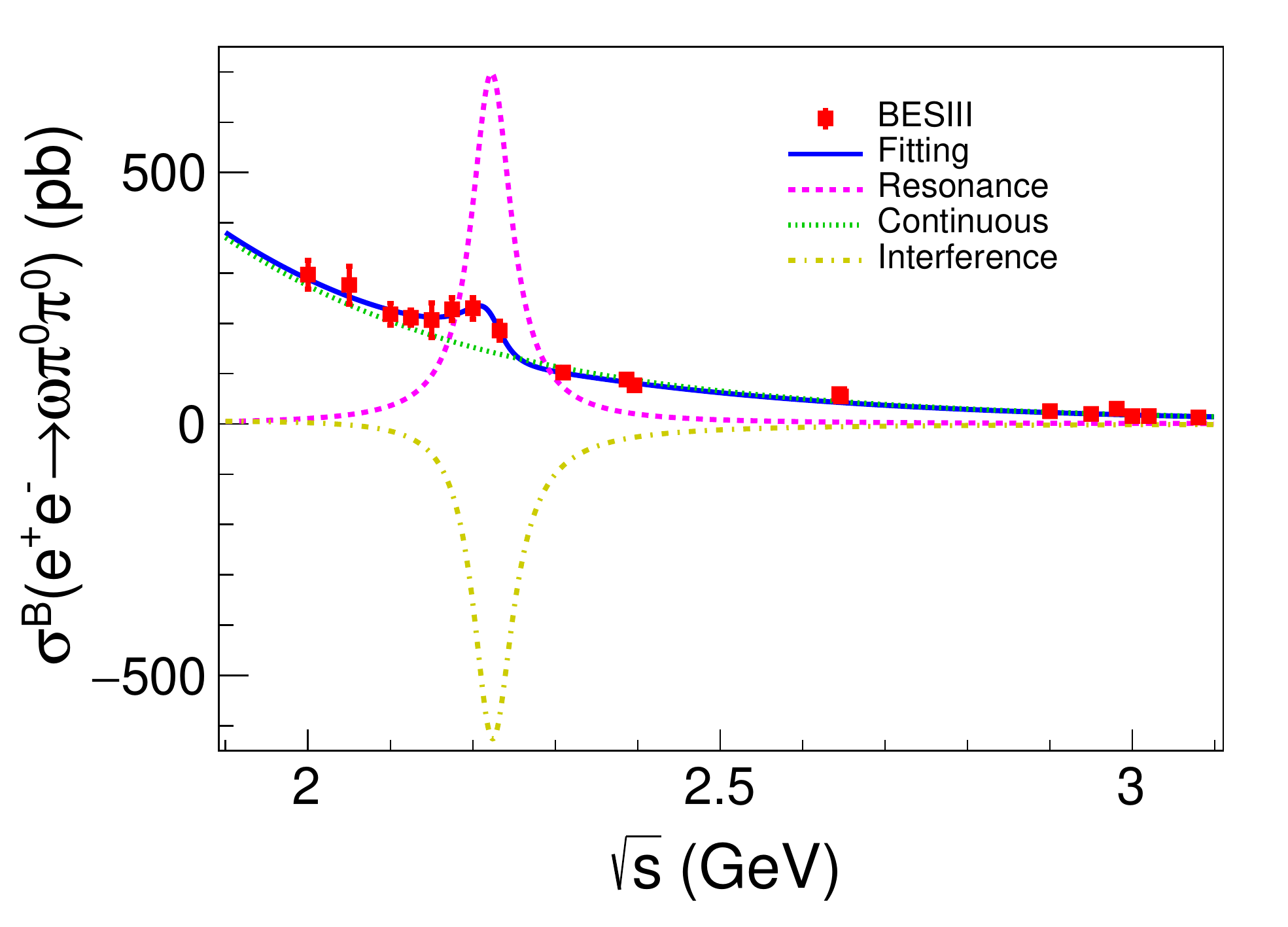}
            \put(25,60){\small\bfseries{(b)}}
        \end{overpic}
    \end{center}
    \caption{Fit to the Born cross section of $\processopp$. (a) The solution with constructive interference. (b) The solution with destructive interference. Red dots with error bars represent data, while blue solid line represents total fitting function, and green dotted, magenta dashed and yellow dashed curves represent the contributions of nonresonant, resonant and interference components, respectively.}
    \label{figure:4:fit}
\end{figure}

\begin{table}[hbpt]
    \begin{center}
        \caption{The result of the fit to the $\processopp$ cross section with functions described by Eq.~(\ref{eq.fitting}). The uncertainty is statistical only. The statistical significance is calculated from the difference in $\chisq$ of the fit if only the continuum shape is used to fit the Born cross sections.}
        \label{table:3:fit}
        \begin{small}
            \begin{tabular}{C{2cm}|R{1cm}PL{1cm}R{1cm}PL{1cm}l}
                \hline\hline
                Parameter                          & \multicolumn{2}{c}{Solution (a)}                                     & \multicolumn{2}{c}{Solution (b)} &              \\
                \hline
                $M_{r}\ ({\rm MeV}/c^2)$           & \multicolumn{4}{c}{$\resultoppmass \pm \resultoppmasserrsta \ \ \ $} &                                                 \\
                $\varGamma_{r}\ ({\rm MeV})$       & \multicolumn{4}{c}{$\resultoppwidth \pm \resultoppwidtherrsta$}      &                                                 \\
                $\phi_{r}$                         & 2.4                                                                  & 0.3                              & -1.7 & 0.1 & \\
                $\varGamma_{ee}^{r}Br\ ({\rm eV})$ & 0.3                                                                  & 0.1                              & 13.8 & 6.6 & \\
                $a(\times 10^3)$                   & \multicolumn{4}{c}{$1.3 \pm 0.2$}                                    &                                                 \\
                $b$                                & \multicolumn{4}{c}{$5.0 \pm 0.1$}                                    &                                                 \\
                Significance                       & \multicolumn{4}{c}{\resultoppcodesignificance}                       &                                                 \\
                \hline\hline
            \end{tabular}
        \end{small}
    \end{center}
\end{table}

To study the systematic uncertainties for the resonant parameters, an alternative fit is carried out by parametrizing the continuum component with an exponential function~\cite{3_1_jpsipipi_1},

\begin{equation}
    \label{eq.alternative}
    \begin{aligned}
         & f_c(s) = \sqrt{\varPhi (\sqrt{s})e^{p_{0}u}p_{1}},
    \end{aligned}
\end{equation} where $p_0$, $p_1$ are floated parameters and $u=\sqrt{s}-(2 M_{\piz}^{\rm PDG}+M_{\omega}^{\rm PDG})$. The differences of the results between the alternative fit with Eq.~(\ref{eq.alternative}) for the continuum component and the nominal fit are taken as systematic uncertainties for the resonant parameters. The systematic uncertainties associated with the signal model are also studied by using a relativistic Breit-Wigner function with an energy-dependent width as the signal shape~\cite{1_PDG}. They are found to be negligible. Finally, the mass and width of the resonance are determined to be \resultoppcodemass\ and \resultoppcodewidth\ with a statistical significance of \resultoppcodesignificance, which is calculated from the change in $\chisq$ of the fit if the resonant contribution is removed. Both statistical and systematic uncertainties of measured Born cross sections are considered in the calculation of $\chisq$ of the fit. In addition, $\varGamma _{e^+e^-}^{r}Br$ is determined to be $0.3\pm0.1\pm0.1~{\rm eV}$ or $13.8\pm6.6\pm5.2~{\rm eV}$ for the two solutions from the fit, where the first uncertainties are statistical and the second ones are systematic.
}

\section{Conclusion}
 {
  The cross section of the process $\processopp$ is measured at nineteen center-of-mass energies from 2.0 to 3.08 GeV with a total integrated luminosity of $\luminosity$. The resonant structure around 2.20~GeV is observed with a statistical significance of \resultoppcodesignificance\ in the coherent fit to the cross section line shape. The resonance has a mass of \resultoppcodemass\ and a width of \resultoppcodewidth, where the first uncertainties are statistical and the second ones are systematic. The resonance observed in this analysis, which could be an $\omega$ excited state, is consistent with \babar's measurement~\cite{3_1_babar_summary}. A future study of this channel with more data sets around $\energy[2.2]$ will be helpful to improve knowledge of this resonance~\cite{4_1_detector_3}.
 }

\section{Acknowledgements}
 {
  The BESIII Collaboration thanks the staff of BEPCII and the IHEP computing center and the supercomputing center of USTC for their strong support. This work is supported in part by National Key Research and Development Program of China under Contracts No.~2020YFA0406400, No.~2020YFA0406300; National Natural Science Foundation of China (NSFC) under Contracts No.~11335008, No.~11625523, No.~11635010, No.~11735014, No.~11822506, No.~11835012, No.~11935015, No.~11935016, No.~11935018, No.~11961141012, No.~12022510, No.~12025502, No.~12035009, No.~12035013, No.~12061131003, No.~11705192, No.~11950410506, No.~12122509, No.~12105276; the Chinese Academy of Sciences (CAS) Large-Scale Scientific Facility Program; Joint Large-Scale Scientific Facility Funds of the NSFC and CAS under Contracts No.~U1732263, No.~U1832207, No.~U1832103, No.~U2032111; CAS Key Research Program of Frontier Sciences under Contract No.~QYZDJ-SSW-SLH040; 100 Talents Program of CAS; INPAC and Shanghai Key Laboratory for Particle Physics and Cosmology; ERC under Contract No.~758462; European Union Horizon 2020 research and innovation programme under Contract No. Marie Sklodowska-Curie Grant Agreement No 894790; German Research Foundation DFG under Contracts No.~443159800, Collaborative Research Center CRC 1044, FOR 2359, FOR 2359, GRK 214; Istituto Nazionale di Fisica Nucleare, Italy; Ministry of Development of Turkey under Contract No. DPT2006K-120470; National Science and Technology fund; Olle Engkvist Foundation under Contract No. 200-0605; STFC (United Kingdom); The Knut and Alice Wallenberg Foundation (Sweden) under Contract No. 2016.0157; The Royal Society, UK under Contracts No.~DH140054, No.~DH160214; The Swedish Research Council; U. S. Department of Energy under Contracts No.~DE-FG02-05ER41374, No.~DE-SC-0012069.
 }

\bibliographystyle{apsrev4-1}
\bibliography{draft.bib}
\end{document}